\documentclass[3p,times,procedia]{elsarticle}
\usepackage{ecrc}
\usepackage{graphicx}
\usepackage{subfigure}

\volume{00}

\firstpage{1}

\journalname{Physics Procedia}

\runauth{Y. Murase et al.}

\jid{phpro}

\usepackage{amssymb}

\biboptions{authoryear}

\usepackage[figuresright]{rotating}

\begin{document}

\begin{frontmatter}



\dochead{27th Annual CSP Workshops on ``Recent Developments in Computer Simulation Studies in Condensed Matter Physics'', CSP 2014}

\title{A tool for parameter-space explorations}


\author[label1,label2]{Yohsuke Murase\corref{cor1}}
\author[label1,label2]{Takeshi Uchitane}
\author[label1,label2,label3]{Nobuyasu Ito}

\address[label1]{RIKEN, Advanced Institute for Computational Science, 7-1-26, Minatojima-minami-machi, Chuo-ku, Kobe, Hyogo, Japan}
\address[label2]{CREST, JST}
\address[label3]{Department of Applied Physics, School of Engineering, The University of Tokyo, 7-3-1 Hongo, Bunkyo-ku, Tokyo 113-8656, Japan}

\begin{abstract}
A software for managing simulation jobs and results, named ``OACIS'', is presented.
It controls a large number of simulation jobs executed in various remote servers,
keeps these results in an organized way, and manages the analyses on these results.
The software has a web browser front end, and users can submit various jobs to appropriate remote hosts from a web browser easily.
After these jobs are finished, all the result files are automatically downloaded from the computational hosts and stored in a traceable way together with the logs of the date, host, and elapsed time of the jobs.
Some visualization functions are also provided so that users can easily grasp the overview of the results distributed in a high-dimensional parameter space.
Thus, OACIS is especially beneficial for the complex simulation models having many parameters for which a lot of parameter searches are required.
By using API of OACIS, it is easy to write a code that automates parameter selection depending on the previous simulation results.
A few examples of the automated parameter selection are also demonstrated.
\end{abstract}

\cortext[cor1]{Corresponding author. Tel.: +81-78-304-4962 ; fax: +81-78-304-4972.}
\end{frontmatter}

\email{yohsuke.murase@riken.jp}



\vspace*{-8pt}
\section{Introduction}
For years, computational power has been increasing exponentially, and it is expected that exa-flops computers are constructed within a decade.
With the advance of computational power and numerical algorithms,
computational sciences is becoming widely applied to various fields and
more and more complex phenomena are becoming targets of computational studies.

As numerical simulations are getting applied to various fields and the models become more complicated,
we usually need to execute a large number of simulation jobs to obtain a meaningful insight.
For example, when we try to reproduce an experimental result by the simulations,
we usually need to run many jobs changing various parameters in order to search reasonable parameters.
As the number of the parameters of a model increases,
the parameter space we need to explore expands exponentially and it becomes infeasible to handle all the jobs manually.
Software which efficiently executes many jobs on distributed hosts and manages the results is strongly expected.

The software is also required to have an interactive user-interface with which we can easily find the results from a bunch of previous simulation results.
This is because we usually repeat the parameter selection depending on the previous results.
Furthermore, in future, such parameter search must be also automated for large-scale parameter-space exploration
because the iteration of the parameter selections can be too large to do it manually.

In this short article, we are going to present a job management application named {\it OACIS},
which stands for {\it Organizing Assistant for Comprehensive and Interactive Simulations}.
By registering a simulator and a host to OACIS, users can easily submit their simulation jobs to the remote host.
Several API are also provided to implement a program for automated parameter search.
A few examples of the automated parameter search are shown in Section \ref{section:auto_param_selection} and \ref{section:applications}.


\section{System Overview}\label{section:system_overview}

\begin{figure}
\begin{center}
\subfigure{
\includegraphics[bb=0 0 222 159, width=.38\textwidth]{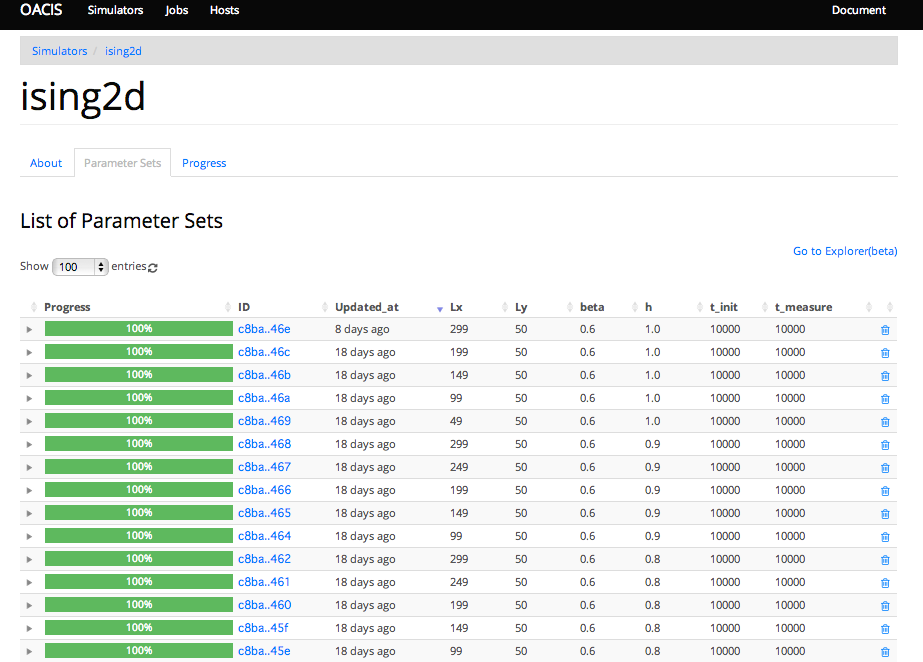}
}
\hspace{.02\textwidth}
\subfigure{
\includegraphics[width=.33\textwidth]{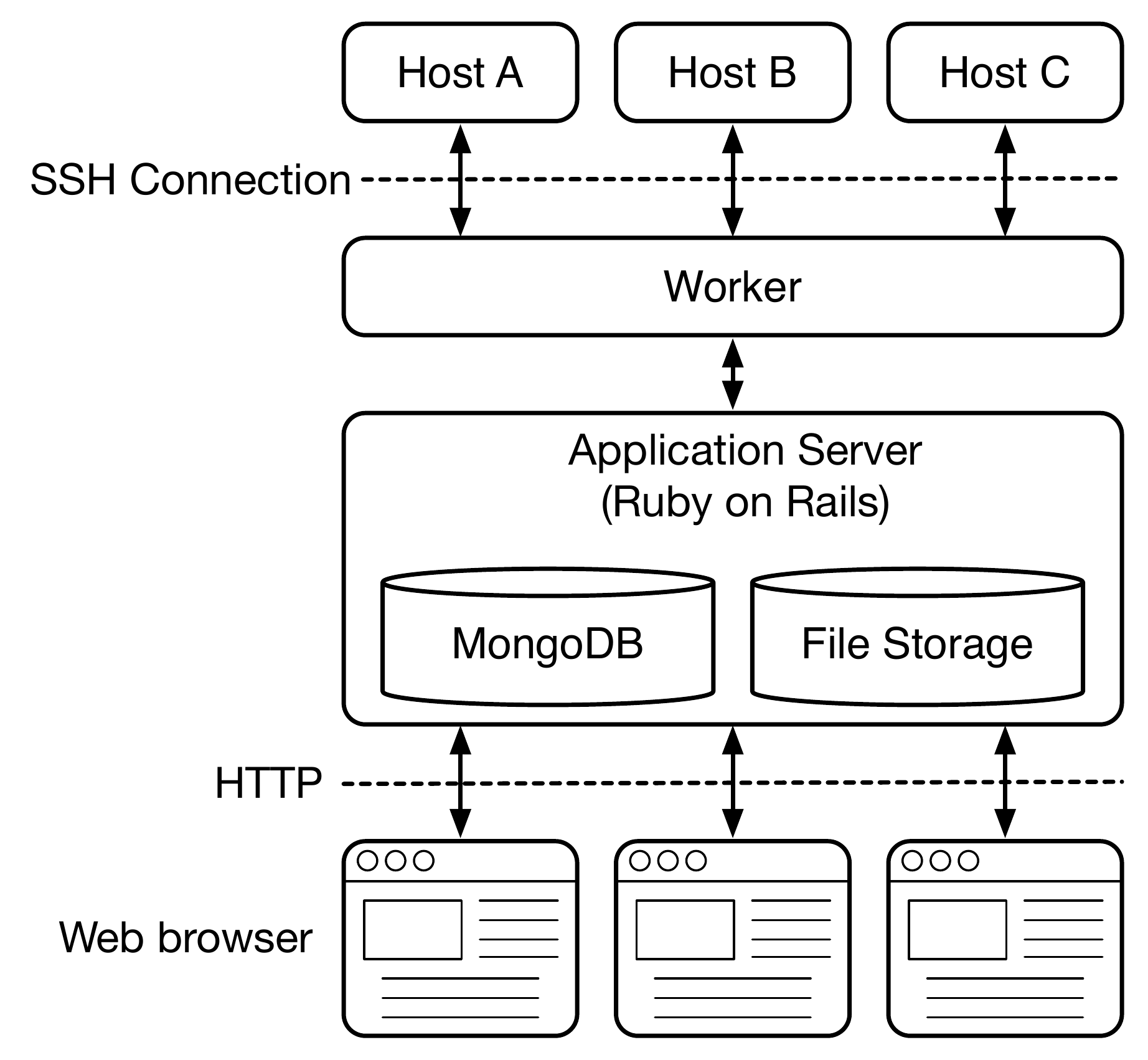}
}
\caption{(left)A snapshot and (right)system overview of OACIS.}
\label{fig:system_overview}
\end{center}
\end{figure}

OACIS has a web browser interface and its snapshot is shown in Fig.~\ref{fig:system_overview}(a).
Once jobs are submitted, OACIS periodically checks if the jobs are finished, and after these jobs finished, OACIS downloads the result files and store these in a file system in a traceable way.
Users can find these simulation results using a web browser.
Not only the simulation results, but also execution logs such as elapsed time, executed host, and the version of the simulation program are also recorded in the database.
Thus, users get free from most of repetitive and error-prone tasks of bookkeeping.

An overview of the system is depicted in Fig.~\ref{fig:system_overview}(b).
The kernel of the system is developed based on Ruby on Rails framework (http://rubyonrails.org/) and internally adopts MongoDB (http://www.mongodb.org/).
This part handles requests from users and manages simulation data.
In addition to this process, there is another daemon process, named ``worker''.
It submits jobs to remote hosts, checks the status of the submitted jobs, and downloads the results after the jobs are finished.

A typical sequence of the job execution is as follows.
When a job is created, the worker checks if there are available hosts to execute this job.
If an available host is found, the worker dynamically generates a shell script to run the registered simulator and submits it to the job scheduler on that host.
Then the worker periodically checks the status of the submitted jobs.
After a job finishes, the worker downloads the result files and store them in a determined place.
Hence, users do not have to check the status of the remote host by themselves.
Furthermore, a job script generated by OACIS is reconfigurable for each host
so that users can assign parameters required by a job scheduler, such as the number of CPUs or the limit of elapsed time.
It is possible to submit jobs either to laptops or high-end super-computers in a unified way.

\begin{figure}
\begin{center}
\includegraphics[bb=0 0 258 195, width=.31\textwidth]{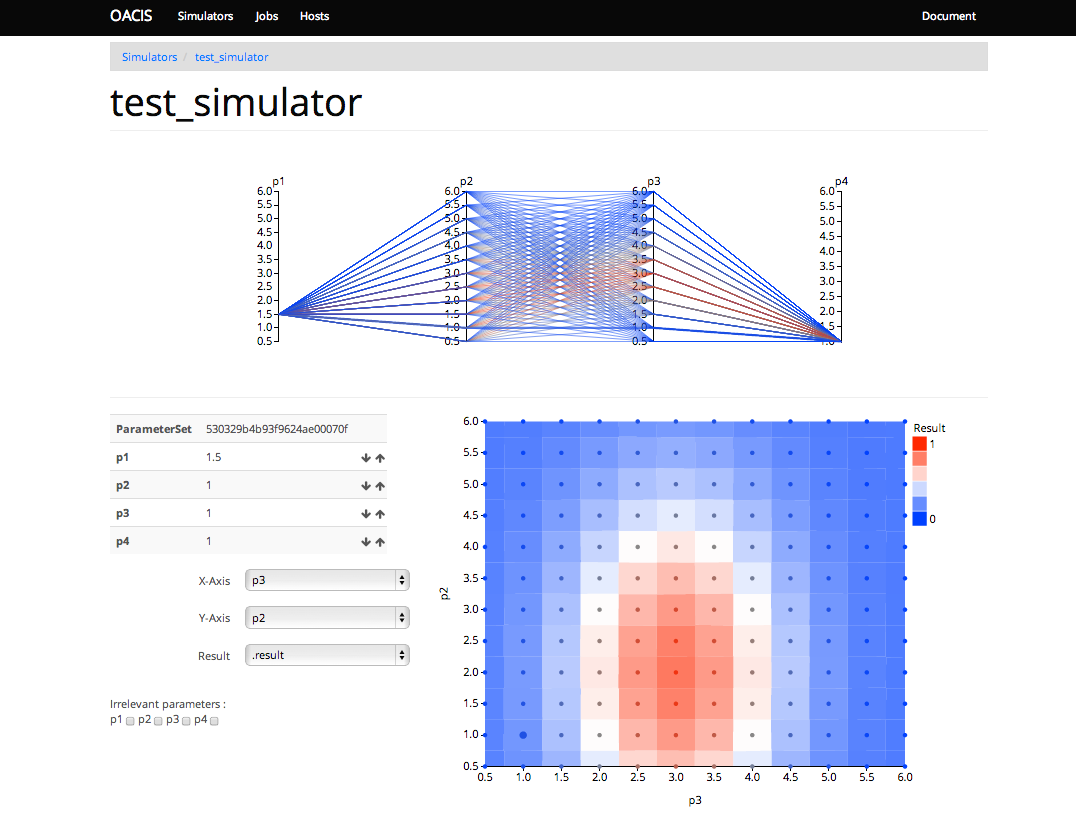}
\caption{A snapshot of a visualization tool in OACIS.}
\label{fig:plot_snapshot}
\end{center}
\end{figure}

Simulation results are stored hierarchically by OACIS.
The database has three layers: ``{\it simulators}'', ``{\it parameter sets}'', and ``{\it runs}''.
The {\it simulator} is a simulation program such as an Ising model simulator or a Potts model simulator.
A {\it parameter set} corresponds to a set of the parameter values.
For example, for the Ising model having two control parameters $\beta$ and $h$,
a set $\{ \beta = 0.5, h = 1.0\}$ corresponds to a parameter set.
A {\it run} is a Monte Carlo run having a unique random number seed.
Each simulator has several parameter sets and each parameter set has several runs.
All the stored results of a run are linked to its simulator and parameter set.

In addition to the functionalities for the job management, some visualization functionalities are implemented.
As shown in Fig.~\ref{fig:plot_snapshot}, results distributing in a parameter space are interactively collected and shown as a plot.
This kind of visualization is useful for researchers to quickly obtain some insights from the results.


\section{Automated Parameter Selection}
\label{section:auto_param_selection}

Although OACIS can handle hundred thousands of jobs,
it is not possible to search the whole parameter space exhaustively
when the number of parameters is not small.
It is necessary to select appropriate parameter sets from all the possible combination of parameters.
OACIS provides Ruby API to automate job creation and analysis of the simulation results.
Using these API, users can write a code that automatically selects parameter sets
based on the previous simulation results.

\begin{figure}
\begin{center}
\subfigure{
\includegraphics[width=.38\textwidth]{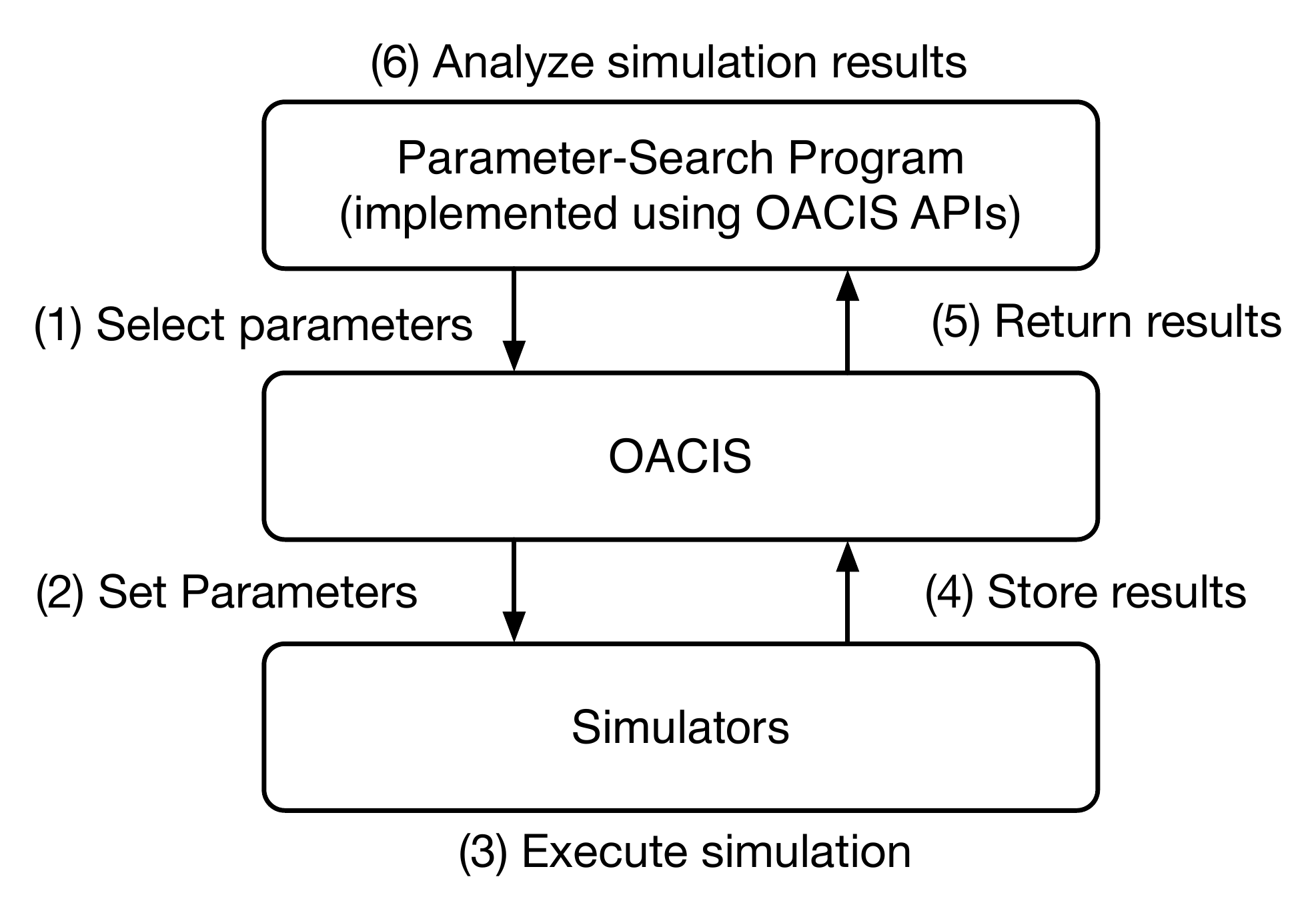}
}
\hspace{.03\textwidth}
\subfigure{
\includegraphics[width=.38\textwidth]{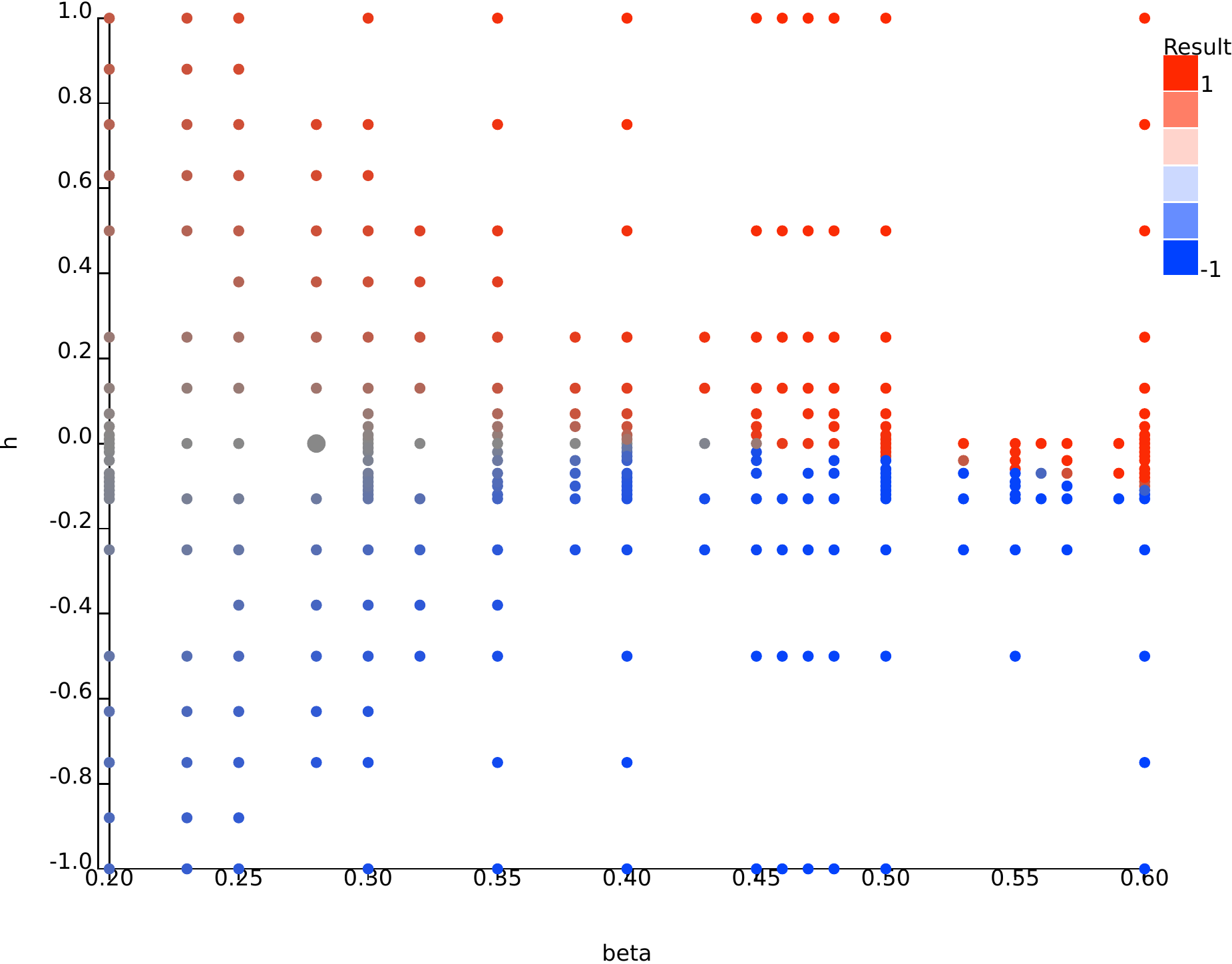}
}
\caption{
(left) A schematic diagram of the system for automated parameter search.
(right) A scatter plot of the parameter sets for the ferromagnetic Ising model on a square lattice.
Each point corresponds to a parameter set where simulation jobs are executed.
Color denotes the order parameter obtained from the simulation.
As you can see in the figure, the area around the first order phase transition is more densely investigated than other area.
}
\label{fig:auto_parameter_search}
\end{center}
\end{figure}

As an example,
we implemented a program to obtain a phase diagram for the Ising model.
Fig.~\ref{fig:auto_parameter_search}(a) shows a schematic diagram of the whole system.
The parameter-search program shown on the top of the figure
selects parameter sets and creates jobs using OACIS API.
OACIS submits these jobs to the remote hosts and the simulator program is executed there.
The simulator shown at the bottom of the figure is a program for the Monte Carlo simulation
for the ferromagnetic Ising model on a square lattice,
which has two parameters $\beta$ and $h$.
The parameter $\beta$ is the inverse temperature and $h$ is the external magnetic field.
The output of the simulator is the averaged order parameter $m$.
After the simulation finishes, OACIS collects the result files and stores them to the database.
The parameter-search program analyzes these results and determine the next parameter sets.
This loop is repeated until sufficient results are obtained.

To obtain the phase diagram efficiently, we used the following algorithm.
Starting from a pre-defined domain in parameter space, $[\beta_{min}, \beta_{max}]$ and $[h_{min}, h_{max}]$,
simulations are executed for each parameter sets corresponding to the vertices of the domain.
If the differences of the obtained order parameters are larger than the threshold value (0.02 is used here), divide the search region into four parts by dividing at the midpoint for $\beta$ and $h$.
For each search region, repeat the above operations until we get sufficient results.
The program was implemented in Ruby using the API functions of OACIS.
The code size is only around 100 lines, indicating that these API functions make the implementation much easier.

Fig.~\ref{fig:auto_parameter_search}(b) shows parameter sets for which simulations are executed.
As shown in the figure, simulations are intensively executed at around $h = 0$ because 
there is a first-order phase transition.
On the other hand, the regions where the results do not show significant dependence on the parameters are sparsely investigated.
(See around $\beta = 0.5$ and $h = 0.8$ for example.)
This dynamic parameter selection contributed to saving the computational and human resources significantly.

\section{Applications to Social Simulations}
\label{section:applications}
OACIS are useful not only for computational physics but also for various computational sciences.
In this section, one of the applications to social simulation is demonstrated.

An optimized signal pattern of urban traffic in a Japanese city (Kashiwa-shi, Chiba-pref.) was searched using a traffic simulator, ``MATES'' (\cite{yoshimura:comp_mod_eng_sci_2006,fujii:j_adv_comp_2011}).
To optimize parameters controlling signal patterns, a parameter-search program was implemented
using API of OACIS in the same way as shown in the previous section.
We employed genetic algorithm to generate new parameter sets with which simulations are executed.
After MATES executes the simulations, the parameter-search program evaluates the results and generates new children, i.e., generates new parameter sets.
This iteration continues until a sufficiently optimized signal pattern is found.
After several iterations, we succeeded in finding a signal pattern that shortens the average travel duration by approximately five percent without exhaustive search over the whole parameter space.

By using OACIS API, the parameter search program does not have to explicitly implement the code for the simulator execution.
When an API is called, OACIS executes simulators at an appropriate host and waits until completion of the simulation.
When the parameter-search program specifies a duplicate parameter set, OACIS returns the existing results immediately.
Thus, OACIS API makes the implementation of the parameter search program much easier.
Furthermore, we can find results of a simulation run later and investigate these in detail because all the results are stored.
Another advantage of using OACIS is that
the parameter search program is reusable for another simulator with a slight modification.
This is because OACIS API hides the interface to the simulator.

\section{Conclusions and Future Perspectives}
In this paper, OACIS and a few examples are briefly presented.
The software is being continuously updated and it will be available as open-source software in a near future.
We expect that it will become possible to investigate quite complex models with OACIS
by automating the interpretation of the simulation results and selection of the parameters.

\section*{Acknowledgments}
We are grateful for helpful comments and advices by I.~Noda and H.~Fujii.






\end{document}